\documentclass[fleqn,usenatbib]{mnras}
\usepackage{newtxtext,newtxmath}
\usepackage[T1]{fontenc}
\usepackage{ae,aecompl}
\usepackage{graphicx}	
\usepackage{amsmath}	
\usepackage{amssymb}	

\title[Magnification in combined shear-clustering analyses]{Disentangling magnification in combined shear-clustering analyses}

\author[Thiele, L., Duncan, C.A.J., Alonso, D.]{
Leander Thiele,$^{1,2,3}$\thanks{lthiele@perimeterinstitute.ca}
Christopher A. J. Duncan,$^{1}$\thanks{christopher.duncan@physics.ox.ac.uk}
David Alonso$^{1}$\thanks{david.alonso@physics.ox.ac.uk}
\\
$^{1}$Astrophysics, Department of Physics, University of Oxford, Denys Wilkinson Building, Keble Road, Oxford, OX1 3RH, UK\\
$^{2}$Perimeter Institute, 31 Caroline Street North, Waterloo, ON, N2L 2Y5, Canada\\
$^{3}$Department of Physics and Astronomy, University of Waterloo, Waterloo, ON, N2L 3G1, Canada\\
}

\date{Accepted XXX. Received YYY; in original form ZZZ}
\pubyear{2019}

\begin{document}
\label{firstpage}
\pagerange{\pageref{firstpage}--\pageref{lastpage}}
\maketitle

\begin{abstract}
We investigate the sensitivity to the effects of lensing magnification on large-scale structure
analyses combining photometric cosmic shear and galaxy clustering data (i.e. the now commonly called
``3$\times$2-point" analysis). 
Using a Fisher matrix bias formalism, we disentangle the contribution to the bias on cosmological parameters caused by ignoring the effects of magnification in a theory fit from individual elements in the data vector, for Stage-III and Stage-IV surveys, assuming well-known redshift distributions, sample selection based on magnitude, and magnification strengths inferred from CFHTLenS and DES Y1 data.
We show that the removal of elements of the data vectors that are dominated by magnification does
not guarantee a reduction in the cosmological bias due to the magnification signal, but can instead
increase the sensitivity to magnification. We find that the most sensitive elements of the data
vector come from the shear-clustering cross-correlations, particularly between the highest-redshift
shear bin and any lower-redshift lens sample, and that the parameters $\Omega_{M}$,
$S_8=\sigma_8\sqrt{\Omega_M/0.3}$ and $w_0$ show the most significant biases for both survey models.
Our forecasts predict that current analyses are not significantly biased by magnification, but this
bias will become highly significant with the continued increase of statistical power in the near
future. We therefore conclude that future surveys should measure and model the magnification as part
of their flagship ``3$\times$2-point" analysis.
\end{abstract}


\section{Introduction}

On-going and future large-scale surveys are in the process of transforming cosmology from a theory- to data-rich field of study. Accordingly, limitations are shifting from statistics to systematics, thus requiring development and implementation of complex models to allow an inference of the underlying cosmological model that is both accurate and precise.

In addition to exquisite measurements of the cosmic microwave background, recent years have seen a surge of precise cosmological measurements from galaxy surveys using weak lensing observations. Such observations are now producing cosmological measurements with a precision comparable to those of the CMB \cite[e.g.][]{KIDS_Joudaki_18, KIDS_Uitert_18,DES_3x2pt_Abbot,DES_Y1_Shear_Troxel,Hildebrandt_KiDS}. With the continuing increase in sky coverage from the Dark Energy Survey (\href{https://www.darkenergysurvey.org}{DES}), the Hyper Suprime-Cam (\href{https://hsc.mtk.nao.ac.jp/ssp/science/weak-lensing-cosmology/}{HSC}) and the Kilo-Degree Survey (\href{http://kids.strw.leidenuniv.nl}{KiDS}), as well as large-area future surveys such as \href{https://www.euclid-ec.org}{Euclid} and the Large Synoptic Survey Telescope (\href{https://www.lsst.org/category/weak-lensing}{LSST}), we can expect that the precision of these measurements will continue to improve.

As a result of this increasing precision, various systematic biases as part of a weak lensing analysis constitute a growing field of study, for example in the context of the lower values of $\sigma_8$ and $\Omega_{\rm M}$ favoured by weak lensing analyses in comparison to Planck \cite[see e.g.][ and other survey-specific analysis papers]{Joudaki_DES+KiDS, DES_Y1_Shear_Troxel}.

The combination of shear observables with clustering data provides a way to internally self-calibrate a range of systematic effects which would be present in a shear or clustering analysis alone. For example, in \cite{JoachimiBridle2010} it was shown that a combination of clustering, cosmic shear and shear-clustering cross-correlations (commonly labelled "galaxy-galaxy-lensing") can improve cosmological parameter constraints by more than the sum of the constituent parts, by lifting degeneracies between the galaxy bias and $\sigma_8$ in the clustering measurements, and shear and intrinsic alignments in the cosmic shear analysis. Such ``3$\times$2-point'' analyses are now routinely undertaken as part of modern weak lensing surveys, using combinations of shear with external spectroscopic data \cite[][]{KIDS_Joudaki_18, KIDS_Uitert_18}, or internal photometric data \citep{DES_3x2pt_Abbot}. Future large scale galaxy surveys such as Euclid and LSST consider photometric 3$\times$2-point analyses as part of their primary science drivers \cite[][]{Euclid_Red_Book,2009arXiv0912.0201L,2018arXiv180901669T}.

Analyses which cross-correlate shear and clustering signals will contain a contribution from the
magnification of the lens and source samples. The clustering signal contains a contribution due to
the correlation between the change in the number density in both the foreground and background,
which are both lensed by any large-scale structure which is foreground of both. The galaxy-galaxy
lensing signal contains a contribution which results from correlations between the magnification of
the lens sample and the shear of the source sample, and a second order contribution from the
magnification of both the lens and source samples \citep{Hui2007,SchmidtRozo2009}. Finally, the shear-shear
signal is unaffected by magnification at first order, but can receive second-order contributions due
to the magnification of the sources and their perturbed observed positions
\citep{shearmag1,SchmidtRozo2009,shearmag2}. The impact of such effects is now well studied, and
recent investigations \citep{Duncanetal2014, Cardona16, Lorenz18, Camera2015} all indicate that whilst the magnification does not significantly contribute to the precision of cosmological constraints, it can catastrophically affect the accuracy of such constraints if it is not appropriately modelled, even where redshift-bin cross-correlations in the clustering data (where magnification is a larger relative contribution to the total signal) are removed from the data vector. Such model bias results from the requirement that the incomplete model (e.g. one which does not include a theoretical prediction for the magnification) must adjust the model parameters to fit data which does include this contribution. Current 3$\times$2-point analyses have not routinely modelled such a contribution in fits to either the clustering or the galaxy-galaxy lensing measurements, but instead the fits are tested for sensitivity to simplified magnification models \mbox{\cite[e.g.][]{DES_Y1_ExtendedModels}}.

In this paper, we use a Fisher matrix bias formalism \citep{2008MNRAS.391..228A} to investigate in depth the sensitivity of each element of the photometric 3$\times$2-point data vector to the magnification bias effect as a means to better understand the origin of the bias due to magnification and its importance in the weak lensing data vector. We focus on the contributions to the bias from individual redshift bin correlations, but we also consider the effect of non-linear scale cuts on the bias. For concreteness, we present two case studies, namely a Stage-III- and a Stage-IV-like survey.

The remainder of this paper is organised as follows: in Section \ref{sec:Theory} we describe the theory models used to describe the shear and clustering observables in a cosmological context, give a description of the surveys we are modelling, and detail the Fisher matrix formalism used for cosmological parameter forecasts; In Section \ref{sec:Results} we present details of the forecast sensitivity of the two analyses considered to the magnification contribution. There, we discuss the novel results of this work: a detailed analysis of the dependence of the total magnification bias effect on individual redshift bin correlations, and the impact of scale cuts on the bias due to magnification. We conclude in Section \ref{sec:Conclusions}. Technical details are given in appendices \ref{app:DESalpha} and \ref{app:contrivedcutting}.

\section{Theory}\label{sec:Theory}
In this section we briefly describe the set of weak lensing observables considered, specify the two surveys we are studying and explain the forecast method. As these contributions are well described elsewhere \cite[e.g.][and many more]{Duncanetal2014, JoachimiBridle2010} we only describe the main results, and the interested reader is encouraged to refer to these works for more detailed information.

\subsection{Weak lensing observables in 3$\times$2-point analyses}
Besides a change in the observed shape of a distant source, gravitational lensing by the large scale structure of the Universe induces a change in the observed position and size (and consequently flux) of the source, collectively referred to as magnification. Direct measures of galaxy size and flux may be used to infer knowledge on cosmology \cite[e.g.][]{Alsing2015} or the lensing mass distribution \cite[e.g.][]{Duncan2016,Schmidt2012} under assumptions on the statistical distribution of intrinsic galaxy properties. Alternatively, the magnification field may also be probed indirectly through measures of galaxy clustering, frequently referred to as `magnification bias'.

Magnification bias is based on the fact that, where a sample of sources is selected on either size or flux, the presence of a local lensing field will induce a deviation in the observed number of sources, as these sources are brought in to (or out of) the sample as a result of the change in these observed properties. Furthermore, the change in the galaxy position induces a change to the local number density due to the dilution (or concentration) of sources in the background of an integrated matter over- (or under-) density. Under the assumption that the distribution of sources as a function of limiting magnitude is well known to the limits of the survey, and the intrinsic distribution of source flux can be well approximated by a power law $N(>S) \propto S^{\alpha}$ near these limits, the local number density at a given angular and radial position on the sky is given by
\begin{align}\nonumber
  n &= n_0(1+\delta_n) = n_0(1+\delta_g + \delta_M + \epsilon) \\
  & \sim  n_0\left(1+b\delta + 2\left(\alpha-1\right)\kappa +\epsilon\right)\,,\label{eqn:N_variation_mag}
\end{align}
where the weak lensing limit has been applied in the final line, position arguments are suppressed
for clarity, we have assumed linear galaxy biasing with bias $b$, and $\delta$ denotes the local
matter density contrast. $\delta_g$ and $\delta_M$ denote the intrinsic galaxy number density
contrast which results from  clustering and magnification due respectively. $\epsilon$ denotes
stochastic variation due to noise. For the remainder of this text, we consider the magnification
contribution only from the application of a selection on magnitude at the faint limit of the survey,
but note that the presence of a bright selection modifies Eqn \ref{eqn:N_variation_mag} such that
$\alpha$ then denotes the difference between the power law index
at the faint and bright end of the survey. Moreover, we do not consider the application of size cuts, but note that the presence of size cuts can induce an extra magnification contribution which can be quantified analogously to the flux magnification contribution in Eqn \ref{eqn:N_variation_mag} if size is quantified as an area \cite[see e.g.][]{SchmidtRozo2009, SchmidtRozo09SizeBias}.

At the faint limit of the survey,
\begin{equation}
  \alpha = 2.5\left.\frac{{\rm d}\log_{10}N(>m)}{{\rm d} m}\right|_{m=m_{\rm lim}}\label{eqn:alpha_m}\,,
\end{equation}
where $m$ corresponds to magnitude.

Under the Limber approximation \citep{Limber1953,Limber1954,Kaiser1992weaklensing}, the projected (angular) power spectrum between redshift bins labelled by indices $i$ and $j$ is given by
\begin{equation}
  C^{ij}_{XY}(\ell) = \int_0^{\chi_H} d\chi\,\frac{W_X^i(\chi)W_Y^j(\chi)}{f_K^2(\chi)}A_X A_Y P_{\delta}\left(k=\frac{\ell}{f_K(\chi)},z(\chi)\right)\,.
\end{equation}
In this equation, $P_\delta(k,z)$ is the matter power spectrum, $X,Y$ labels either shear ($G$) or each constituent element of the number density fluctuations ($g,M$, we label the total contribution to the number count fluctuation $n$), $W$ gives the radial weight function, $A_X$ the prefactor which links the measurable $X$ to the matter density contrast, and $\ell$ is the angular wavenumber. We distinguish the following cases (analogously for $Y$):
\begin{enumerate}
  \item Where $X=G$ labels the shear, $W_X$ denotes the lensing kernel (see Eq. 3 of \cite{Duncanetal2014}), and $A_X = 1$.
  \item Where $X=g$ labels the intrinsic clustering, $W_X$ denotes the galaxy redshift distribution and $A_X=b$, the galaxy bias.
  \item Where $X=M$ labels the magnification bias contribution, $W_X$ is the lensing efficiency and $A_X=2(\alpha-1)\kappa$.
\end{enumerate}

In the presence of intrinsic clustering alongside magnification, the projected position-position power spectra will be comprised of four contributions:
\begin{equation}
    C_{nn}^{ij} = (C_{gg} + C_{gM} + C_{Mg} + C_{MM})^{ij}\,,
\end{equation}
where the dependence on angular wavenumber $\ell$ is implied. The first of the contributions corresponds to the intrinsic clustering signal, which describes correlations due to sources in the two samples clustering as a result of experiencing the same local gravitational potential. This first term is zero where the both bins are widely separated (so there is no overlap between $W^i$ and $W^j$). Assuming the $i$-th bin is in the foreground of the $j$-th bin, the second term denotes the correlation due to the magnification of the background sample by the matter environment local to the foreground. The third term gives the magnification of the foreground due to the local matter environment of the background, and is zero except where there is a non-zero probability that the background exists in front of the foreground (i.e. when $W^i$ and $W^j$ overlap), as can be the case in the presence of photometric redshift uncertainty. The final term gives the correlations due to the magnification of both the foreground and background by the large scale structure between the foreground and the observer. As the lensing kernel is broad, the magnification terms are typically non-zero for all redshift combinations, and dominate over the intrinsic clustering term only where both samples are widely separated in redshift.

Similarly, the position-shear (or galaxy-galaxy lensing) contribution follows
\citep{Hilbert2009, ZiourHui2008}
\begin{equation}
    C_{nG}^{ij} = (C_{gG} + C_{MG})^{ij}.
\end{equation}
The first term is the standard galaxy-galaxy lensing signal, describing the correlation due to the shearing of the background sample by the foreground local matter environment. The second term gives the correlation which results from the  magnification of the foreground and shearing of the background due to common large scale structure between the foreground and observer.

We note that in this section (and in the remainder of this paper)
we are considering only the contribution of the magnification to the local number density, and as a result there is no magnification contribution to the shear-shear correlations $C_{GG}$. However, the shear signal can also display non-zero variation due to the magnification effect. For example, the shear sample will also be selected by flux or size (e.g. through signal-to-noise considerations or in star-galaxy separation) which can induce magnification-dependent correlations \cite[see][]{SchmidtRozo2009,Liu2014}. Moreover, the accuracy with which galaxy shear can be measured is also signal-to-noise or size dependent which can induce further magnification-dependent correlations through the shear multiplicative and additive biases. These second-order effects \citep{shearmag1,shearmag2} will give an additional contribution to the position-shear and shear-shear power spectra, and their importance will be investigated in future work.

\begin{figure*}
\includegraphics[width=\textwidth]{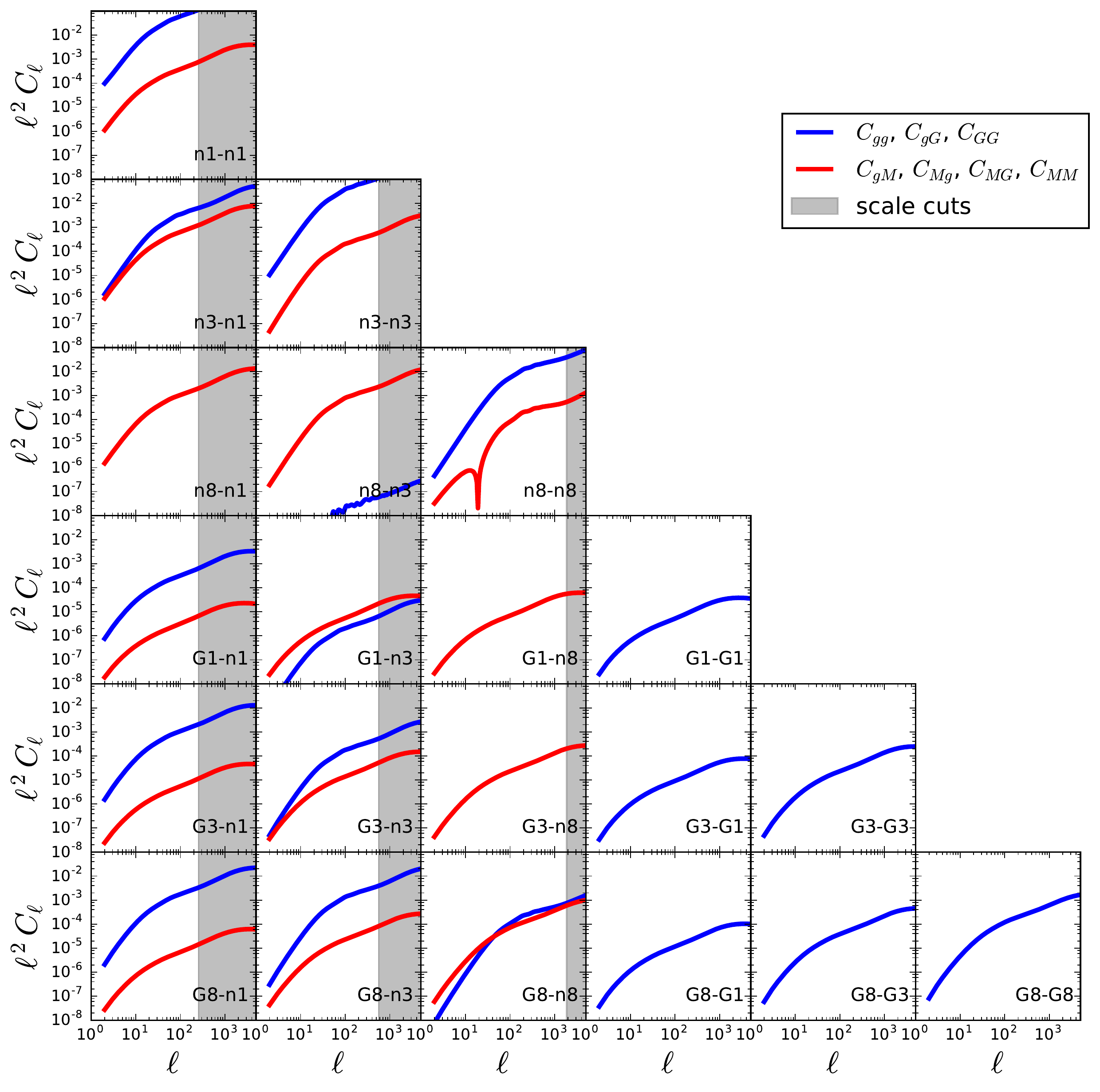}
\caption{The contribution of the magnification signal to the $C_{\ell}$s, for the Euclid-like survey. Selected redshift bin combinations only are shown. \textbf{n} stands for clustering (position measurements), \textbf{G} for shear. We plot the signal without the magnification contribution in solid blue, the absolute value of the magnification contribution is plotted in orange. The scale cuts employed for our Euclid-like setup are indicated by the shaded regions.}\label{fig:GGL-power-spectra}
\end{figure*}

To understand the behaviour of each of the various contributions described to the position-position and position-shear power spectra, the reader is encouraged to see Fig.~\ref{fig:GGL-power-spectra} \cite[compare Fig. 1 of][]{JoachimiBridle2010}. The power spectra plotted there are computed for the fiducial Stage IV Euclid-like survey, which is described in Sec.~\ref{sec:SurveyModel_Euclid}. In the case of the position-shear power spectra, scales on which the magnification term becomes dominant are less simple compared to the position-position spectra. However, we can see that generally the magnification becomes a larger contribution to the total power in shear-position cross-correlations when 
\begin{enumerate}
    \item both lens and source redshift bins increase,\label{en:ggl_item1}
    \item the lens bin redshift increases for a fixed source bin,\label{en:ggl_item2}
    \item the source bin decreases for a fixed lens bin, and\label{en:ggl_item3}
    \item the source sample has a larger probability of existing in front of the lens sample.\label{en:ggl_item4}
\end{enumerate}
Case \ref{en:ggl_item1} occurs because as both samples increase in redshift, both are magnified by a
larger proportion of large scale structure along their common line-of-sight, compared to the matter
environment local to the foreground. Similar arguments explain \ref{en:ggl_item2} and
\ref{en:ggl_item3}. In case \ref{en:ggl_item4}, where the source sample is in front of the lens sample, the local environment of the lens sample cannot contribute to the shearing of the source sample, and any correlation exists solely as a result of the large scale structure along their common line-of-sight.

\subsection{Fisher matrix constraints and bias forecasts}
We produce parameter forecasts by constructing a Fisher matrix as
\begin{equation}
  F_{\eta \tau} = \sum_{\ell} \partial_\eta {\bf C}_{\ell}{\sf Cov}^{-1}_\ell\partial_{\tau}{\bf C}_{\ell}
\end{equation}
where ${\bf C}_{\ell}$ is the data vector constructed of the position-position, position-shear and shear-shear power spectrum across all redshift bin combinations, ${\partial_\tau}$ denotes the derivative of the data vector with respect to parameter $\theta_\tau$ around the fiducial cosmology, and ${\sf Cov}$ is the covariance matrix for this data vector \cite[for further detail, including construction of the covariance matrix and ordering of the data vector, refer to ][]{Duncanetal2014}. The Fisher matrix is constructed utilising {\tt GoFish}\footnote{\url{https://github.com/damonge/GoFish}}, and the power spectra are produced by {\tt CLASS} \citep{Lesgourgues2011,DiDio2013}. In the remainder of the text, greek indices will refer to cosmological parameters, whilst capital roman indices will refer to redshift bin {\sl pairs}.

From the Fisher matrix, we determine the bias on cosmological parameter $\theta_\eta$ as \cite{2008MNRAS.391..228A}
\begin{equation}
  \Delta \theta_\eta = \sum_I \Delta \theta^I_\eta,
\end{equation}
with
\begin{eqnarray}
  \Delta \theta^I_\eta &=&\sum_{\ell} \sum_{J}\sum_{\tau} (F^{-1})_{\eta \tau}\partial_{\tau}C^J_{\ell} ({\sf Cov}^{-1})^{J,I}_{\ell}\Delta_{\alpha} C^I_\ell \label{eqn:FMBiasZ}\\
  &=& \sum_{\tau} (F^{-1})_{\eta \tau}\mathcal{F}^I_{\tau\alpha }\Delta \alpha
  \label{eqn:FMBiasZlin}\,,
\end{eqnarray}
where the redshift bin combinations have been made explicit, $I$ and $J$ refer to redshift bin pairs and $\Delta_\alpha{\bf C}$ refers to the change in the data vector as a result of a change in the magnification parameter $\alpha$. In this analysis, we consider $\Delta_\alpha{\bf C}$ as the difference between the case with magnification ($\alpha \ne 1$) to the model without magnification (equivalent to $\alpha = 1$). In the second line we have approximated the change in the power spectrum to linear order in the change in $\alpha$ ($\Delta_\alpha {\bf C}_\ell\simeq \Delta\alpha\,\partial_\alpha {\bf C}_\ell$), and defined the pseudo-Fisher matrix
\begin{equation}
  \mathcal{F}^I_{\alpha\eta}\equiv \sum_\ell\sum_J \partial_\alpha C^I_\ell\,({\sf Cov}^{-1})^{I,J}_\ell \partial_\eta C^J_\ell.
\end{equation}
$\mathcal{F}_{\alpha \eta}$ describes the (inverse) covariance between the magnification parameter $\alpha$ and the free parameter $\theta_\tau$. The contribution to the bias due to a systematic change in fixed parameter $\alpha$ on free parameter $\eta$ therefore depends on the interplay of the magnitude of their covariance (given by $\mathcal{F}$) and the covariance between different free parameters, encoded in $F$.

Notably, Eq.~\ref{eqn:FMBiasZ} allows us to consider the contribution to the total parameter bias from each element of the data vector, and therefore a means to distinguish those elements of the data vector which are most affected by the magnification bias effect in terms of final constraints, as well as how all elements combine to give the overall bias. We note that the dependence of the cosmological bias on the fiducial $\alpha$ can be inferred from Eq.~\ref{eqn:FMBiasZlin} for small perturbations around the fiducial value\footnote{Note that formally this holds only for small perturbations, however since the dominant magnification contribution to the position-position signal is $C_{Mg} + C_{gM} \propto \alpha-1$ and the magnification contribution to the position-shear signal is linear in $\alpha-1$, the linearisation of Eq.~\ref{eqn:FMBiasZlin} is roughly applicable to larger
perturbations provided that the foreground and background samples are described by the same
$\alpha$.}, although the determination of cosmological bias in this analysis only uses the form in
Eq.~\ref{eqn:FMBiasZ}.

\subsection{Survey modelling}
To investigate the importance of the magnification bias in the data vector elements of a photometric 3x2pt analysis, we consider two main case studies: in the first, we consider a future Euclid-like survey, characterised as a survey with a high number density and accurate photometric redshifts. This case study is aimed at extending the results of \cite{Duncanetal2014} and \cite{Lorenz18}, and in particular to disentangle the biggest contributions to the parameter bias in the data vector. In the second scenario, we consider a current survey, mimicking the 3$\times$2-point analysis of the DES data \citep{DES_3x2pt_Abbot}, and aimed at probing the potential sensitivity to magnification bias in this analysis. In this case, we utilise the publicly available DES Year-1 (Y1) catalogue \citep{DES_Y1_DR1_Cite, DES_Y1_ImageProcess_Cite, DES_Y1_Camera_Cite} to characterise the magnification bias contribution in the current analysis, and to guide the determination for the full survey.  Further details of each case are given below.

Although the data is available, we do not consider the Kilo-Degree survey analyses \citep{KIDS_Joudaki_18,KIDS_Uitert_18} as an example, as the overlapping area used for the clustering and galaxy-galaxy lensing signal is smaller than the area used for the cosmic shear analysis. As a result, we expect the sensitivity to magnification in the KiDS analyses to be smaller than that of DES, where the clustering signal is constructed over the full survey area. We note however that the magnification effect will still be present in the KiDS data vector, even where the clustering is measured using spectroscopic redshifts, and particularly in the galaxy-galaxy lensing analysis.

In this work, we consider as fiducial cosmology a $w$CDM model with seven free cosmological
parameters: the matter density parameter $\Omega_{\rm M} = 0.316$, the baryon density parameter
$\Omega_{\rm B} = 0.049$, the amplitude of linear matter perturbations $S_8 =
\sigma_8\sqrt{\Omega_{\rm M}/0.3} = 0.8$, the dimensionless Hubble constant $h = 0.69$, the scalar
spectral index $n_s = 0.966$, the curvature parameter $\Omega_{\rm K} = 0$, and the Dark Energy
equation-of-state $w_0 = -1$.

\subsubsection{A Euclid-like (Stage IV) survey}\label{sec:SurveyModel_Euclid}
The Euclid-like survey closely follows the set-up of the Stage-IV type survey in \cite{Duncanetal2014}. The true galaxy redshift distribution is modelled as in \cite{SmailEllisFitchett1994}, normalised to give $28$ galaxies per square arcminute. Photometric redshift uncertainties are modelled as a convolution with a Gaussian kernel, with width $\sigma_z = 0.05(1+z_{\rm phot})$.

Galaxy shape noise variance is taken as $\sigma_{\gamma}^2=\sigma_\epsilon^2/2$, where $\sigma_\epsilon = 0.4$ is the total intrinsic ellipticity standard deviation. For the clustering signal, we cut non-linear scales by applying angular frequency cuts at $\ell_{\rm max}(z) = f_K[\chi(z)]k_{\rm max}(z)$ as a function of redshift bin $z$, where $f_K(\chi)$ is the comoving angular distance, $k_{\rm max}$ is determined as in \cite{Duncanetal2014}: $k_{\rm max} = 1.4\pi R_{\rm min}^{-1}$, and the critical scale $R_{\rm min}$ is determined by requiring that the variations in the matter overdensity smoothed over that scale do not exceed a chosen value:
\begin{equation}\label{equ:sigmaRdef}
  \sigma_R^2 = \int d\log k\,\Delta^2(k,z) W^2(k R_{\rm min}(z))\,.
\end{equation}
Here, the window function is chosen as a spherical top-hat in real space (i.e. $W(x) = 3x^{-3}(\sin
x - x\cos x)$ in Fourier space). Our fiducial scale cut is then defined by setting $\sigma_R=0.5$,
in agreement with the choice made in \cite{Duncanetal2014}. As in previous work, we choose the median redshift in each redshift bin as the point where to evaluate $\ell_{\rm max}(z)$. We consider the impact of these scale cuts on both constraining power and magnification bias in more detail in Sec.~\ref{sec:scalecuts}. We model galaxy bias as linear, with $b_{\rm fid} = 1$ for all redshift bins. However, we allow one free galaxy bias parameter in each redshift bin, so that it is determined simultaneously with the cosmological parameters. For the shear power spectra, we set $\ell_{\rm max} = 5000$. Sky coverage is $15,000$ sq-degrees, corresponding to $f_{\rm sky} = 0.37$.

\begin{figure}
\includegraphics[width=0.5\textwidth]{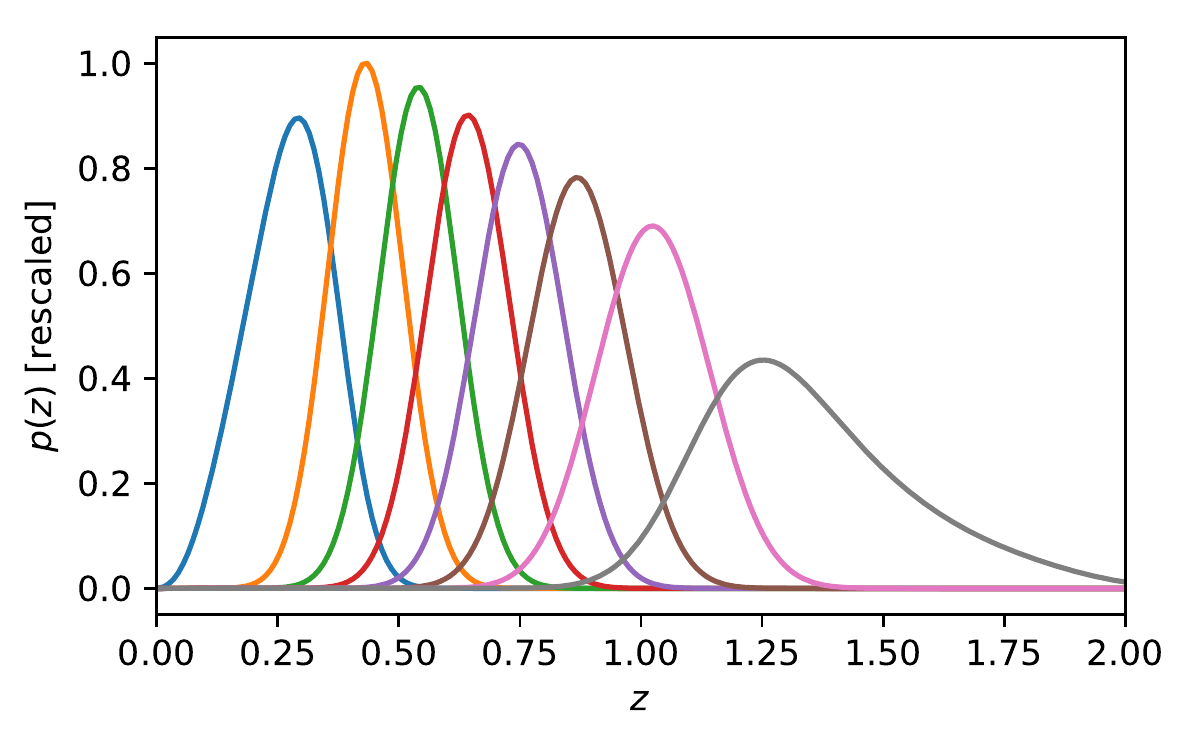}
\caption{Probability distribution of galaxies used in our model of a Stage IV survey. The ordinate units are arbitrary.}\label{fig:ndensities}
\end{figure}
We produce results for 8 tomographic redshift bins, which are used as both foreground and background
samples for all observables. This number is chosen to aid the visualisation of the contribution to
the parameter bias as a function of redshift bin correlations for each observable across the data
vector. The assumed distribution of galaxies in these 8 bins as a function of redshift is plotted in
Fig.~\ref{fig:ndensities}. The magnification strength is $\alpha = 0.7$ for all redshift bins,
corresponding to the fiducial case of \cite{Duncanetal2014} (Section 3.2) which was inferred from
the Canada-France-Hawaii-Telescope Lensing survey (\href{http://www.cfhtlens.org}{CFHTLenS}) catalog. 

\subsubsection{A DES-like (Stage III) survey} \label{sec:SurveyModel_DES}
In modelling a DES-like survey, we closely follow the Y1 data release. We take the source and lens number densities according to the Y1 results, which measured galaxy shapes in 4 redshift bins and number densities in 5 redshift bins \cite[see Fig 1 of][for the redshift distributions used]{DES_3x2pt_Abbot}. The scale cuts follow the DES Y1 prescription, but for simplicity we cut both the position-position and the position-shear power spectra at the same scale (co-moving $8\,h^{-1}{\rm Mpc}$). We then translate angular separations into conjugate space according to $\ell_{\rm max} = N/\theta_{\rm min}$, with $N=2$. This value of $N$ was chosen in order to match the constraining power found in the DES Y1 analysis. Since the DES analysis used real-space correlation functions, it is not possible to draw a one-to-one correspondence between their scale cuts and our cuts imposed directly in harmonic space. We take the fiducial value for redshift-dependent galaxy bias according to \cite{DES_2pt_Clustering_ElvinPoole}, but allow it to vary independently in each redshift bin. Galaxy shape noise variance is modelled as for the Euclid-like setup. The fiducial values for the magnification strength in each clustering redshift bin, $\alpha = \{2.0, 1.5, 0.3, 0.3, 0.3\}$, are inferred from DES Y1 data, according to the procedure described in Appendix~\ref{app:DESalpha}. In our fiducial results we apply cuts to the data vector in accord with the DES Y1 analysis, removing position-position cross-correlations (but keeping the auto-correlations).

We provide results for two different values of sky coverage. The first version with $f_{\rm sky} = 0.032$  (or $1321\,{\rm deg}^2$) is analogous to the Y1 survey, while the second version with $f_{\rm sky} = 0.121$ (or $5000\,{\rm deg}^2$) follows the projected full DES data products. We note that the second version is expected to be conservative relative to the final data quality, since we do not account for the increase in the number of high-redshift sources relative to the Y1 data nor any other improvements in data quality or analysis technique.

\section{Results}\label{sec:Results}
Having described both the formalism used to estimate parameter biases from magnification at Fisher matrix level as well as the assumptions we use to model the two survey types, we can now proceed to presenting our results. First, we will discuss the parameter biases computed in our fiducial setups. Second, we will consider the total parameter bias from magnification as a sum of biases arising from individual redshift bin combinations (c.f. Eq.~\ref{eqn:FMBiasZ}). Finally, we will discuss the impact of scale cuts (which are usually intended to mitigate biases from poorly understood small scale physics) on the magnification bias.

\subsection{Cosmological parameter bias from magnification}
\begin{table}
\begin{tabular}{cccc}
		& DES Y1			& DES Full		& Euclid		 	\\
\hline                                                                                                  
Parameter	& $\Delta/\sigma$		& $\Delta/\sigma$	& $\Delta/\sigma$		\\
\hline                                                                                                  
$\Omega_{\rm M}$& -1.4		    		&	-2.73		& 8.8           		\\
$\Omega_{\rm B}$& -0.032            		&	-0.064		& -0.29           		\\
$S_8$           & -0.91		      		&	-1.77		& 6.5           		\\
$h$             & 0.53            		&	1.03		& -2.8            		\\
$n_s$           & -0.13            		&	-0.25		& -3.2            		\\
$\Omega_{\rm K}$& -0.22           		&       -0.43			& 2.0            		\\
$w_0$           & -1.2		      		&	-2.34		& 4.4            		\\
\hline                                                                                                  
relative FoM	& 1 (per def.)			&	3.8		& 	115			\\
\hline
\end{tabular}
\caption{Ratio of the biases on cosmological parameters to their standard errors, for the three survey specifications we are considering in this work. DES Y1 has sky coverage $f_{\rm sky}=0.032$, while DES Full has $f_{\rm sky}=0.121$. The Euclid-like sample is described in Section \ref{sec:SurveyModel_Euclid}. The last line lists the relative all-parameter figure of merit, normalised to 1 for DES Y1.}
\label{tab:biasesDESEuclid}
\end{table}

We validated our Fisher matrix by comparing the predicted parameter uncertainties to
\cite{Duncanetal2014} for the Euclid-like case; and against the measured parameter uncertainties of
\cite{DES_3x2pt_Abbot} for the DES-Y1-like case, with good agreement to between $<50\,\%$ for Euclid and
$<30\,\%$ for DES. In both cases we note that we utilise a different cosmological parameter set. We emphasise that for the DES-like case we are comparing the Fisher expected parameter uncertainty to the measured uncertainty. Moreover, as we do not model all the complicating factors of the DES model (e.g. nuisance parameters on redshift distribution mean as an example), and we work in Fourier space, whereas the DES results were derived in configuration space, we consider this a good enough agreement to describe the main aspects of the analysis.

In Tab.~\ref{tab:biasesDESEuclid} we list the ratio of bias $\Delta$ to standard error $\sigma$ on seven cosmological parameters, as well as the relative figure of merit (FoM) for two versions of the DES survey (which differ only by sky coverage) and the Euclid survey. We define the FoM as
\begin{equation}
  \text{FoM} = \left(\det(F^{-1})_p\right)^{-1/n_p},
\end{equation}
where $p$ is the set of cosmological parameters (i.e. excluding the galaxy bias parameters we marginalise over) and $n_p$ is the number of parameters in the set.

For the Euclid-like survey, we find that with the sole exception of $\Omega_{\rm B}$, all parameters are significantly biased by the neglect of the magnification signal as found in \cite{Duncanetal2014,Lorenz18,Cardona16} and reinforcing the conclusion that magnification must be measured and modelled to avoid catastrophic parameter bias in the 3x2pt analysis of Euclid data. 

\begin{figure*}
  \includegraphics[width=\textwidth]{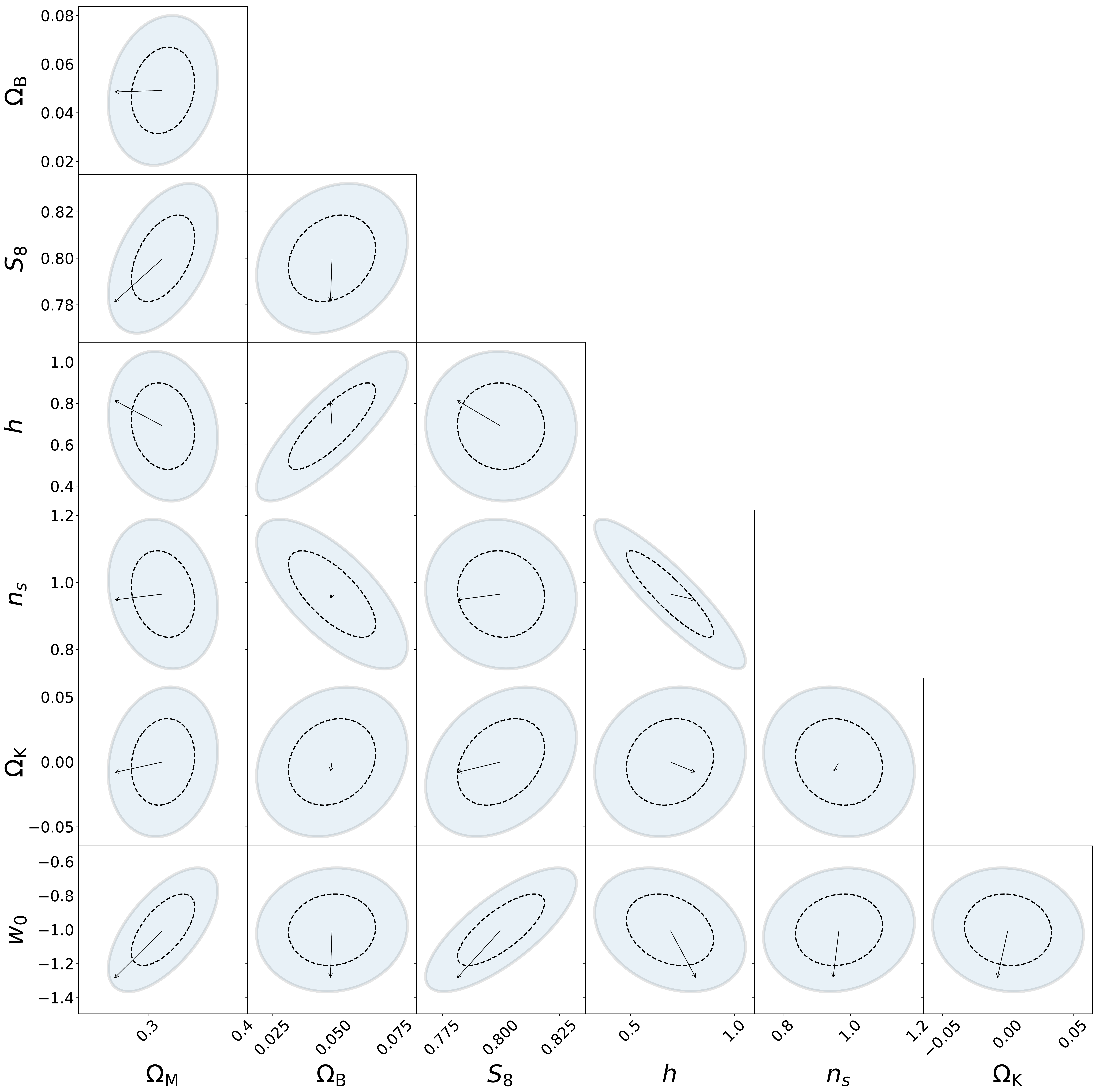}
  \caption{1$\sigma$ confidence regions for the cosmological parameters considered here, with the estimated Fisher biases indicated by arrows. The shaded ellipses correspond to the DES Y1 results, while the dashed ellipses are for our simple model of a full DES survey (which is only increased in sky coverage in comparison to Y1). While the estimated biases on the Y1 set-up are smaller than the statistical uncertainty on the inferred parameters, the increase in statistical power resulting from the extension to the full DES sky coverage causes the bias in some parameters to become significant.}
\label{fig:ellipsesDES}
\end{figure*}

For the case of the DES-Y1-like set-up, we find that $S_8$, $\Omega_{\rm M}$ and $w_0$ are the most affected by magnification, with a predicted bias of comparable size to their statistical uncertainty. Where the survey area has been increased to the full DES sky coverage, we find that these parameters become significantly biased as a result of the increased statistical power of the survey. These results are also summarised in Fig.~\ref{fig:ellipsesDES}. This suggests that current surveys are now reaching the point where the contribution of the magnification must be modelled. 

In interpreting these results, we reiterate that whilst the model used for the DES-Y1-like survey mimics the main effects of that analysis, the full analysis of the DES-Y1 results contains many extra contributions which are not modelled here.
Moreover, we emphasise that
the lower values of $S_8$ currently seen in DES weak lensing analyses compared to Planck
CMB inferred values are also seen in the analysis of shear-shear correlations only
\citep{DES_Y1_Shear_Troxel}.
Whilst magnification can 
impact the shear-shear signal through source sample selection
\citep[see Section~\ref{sec:Conclusions} for further
discussion, or, e.g.,][]{SchmidtRozo2009},
this is not modelled in this analysis, and therefore
the observed parameter shift cannot be sourced
by the terms we investigate here. 
Further, we note that the Fisher matrix approach will tend to underestimate statistical uncertainty (as it satisfies the Cram\'{e}r-Rao bound), while overpredicting the bias in non-Gaussian posteriors. As such the results presented should be interpreted as indicative of the sensitivity of the survey to the magnification, but cannot be used to accurately infer absolute biases.

\subsection{Disentangling the contributions from redshift bin correlations}

\begin{table*}
\begin{tabular}{cccc}
	       & no cuts		& --clustering cross-correlations & --GGL shear foreground	\\
\hline
Parameter	& $\Delta/\sigma$	& $\Delta/\sigma$		& $\Delta/\sigma$		\\
\hline
$\Omega_{\rm M}$& 8.8 			& 11				& 12				\\
$\Omega_{\rm B}$& -0.29			& 0.98				& 0.78				\\
$S_8$           & 6.5			& 7.2				& 7.2				\\
$h$             & -2.8			& -2.7				& -4.1				\\
$n_s$           & -3.2			& -4.0				& -3.4				\\
$\Omega_{\rm K}$& 2.0			& 1.2				& 1.3				\\
$w_0$           & 4.4			& 6.4				& 7.0				\\
\hline
relative FoM	& 1 (per def.)		& 0.81				& 0.79				\\
\end{tabular}
\caption{Effect of two different cutting schemes on the relative bias and the all-parameter figure of merit, compared to our fiducial setup where no cuts are employed. These results are for the Euclid-like survey model.
In the second column (--clustering cross-correlations), we remove cross-correlations between
clustering power spectra in different redshift bins in the Fisher matrix. In the third column (--GGL
shear foreground), we additionally remove those parts of the galaxy-galaxy lensing signal stemming
from power spectra using the shear sample in a foreground redshift bin.}\label{tab:Euclidcuttingbiases}
\end{table*}

We now proceed in building up intuition on the source of the magnification bias -- first, by looking at different redshift bin correlations individually.

Referring back to the discussion in Sec.~\ref{sec:Theory} and the power spectra in Fig.~\ref{fig:GGL-power-spectra}, a dominant magnification contribution may be expected primarily for cross-correlations of two widely separated clustering bins and correlations between a foreground shear bin and background clustering bin. We now investigate the effect of removing these redshift bin combinations from the analysis of Stage-IV-like data (in fact, as discussed before, the DES Y1 analysis already removed clustering cross-correlations, while keeping the magnification dominated ``inverted'' galaxy-galaxy lensing correlations in order to self-calibrate the data).

In Tab.~\ref{tab:Euclidcuttingbiases}, we present relative biases and FoMs for the Euclid model and three different redshift bin cutting schemes. In the left column, we list our fiducial results without application of any cuts (these have already been given in Tab.~\ref{tab:biasesDESEuclid}). In the centre column, we remove clustering cross-correlations, while in the third column we also cut out correlations between shear foreground and clustering background bins. From the relative FoMs listed, it can be seen that the clustering cross-correlations contributed about $20\,\%$ of the total constraining power, while the bin combinations removed in the last step add only negligible information (which should not be surprising since these bins are entirely dominated by the magnification signal, which is sub-dominant in determining the cosmological precision in $3\times 2$-point analyses). Perhaps more surprisingly, we find the biases not to behave as intuitively expected. In fact, removing redshift bin correlations that we know to be substantially contaminated by the magnification signal does not necessarily decrease the relative bias. We note that this agrees with \cite{Duncanetal2014} (Table 2), which also found that removing clustering cross-correlations in a 3x2pt analysis can increase sensitivity to magnification whilst decreasing the FoM by ~$20\%$.

\begin{figure*}
\includegraphics[width=\textwidth]{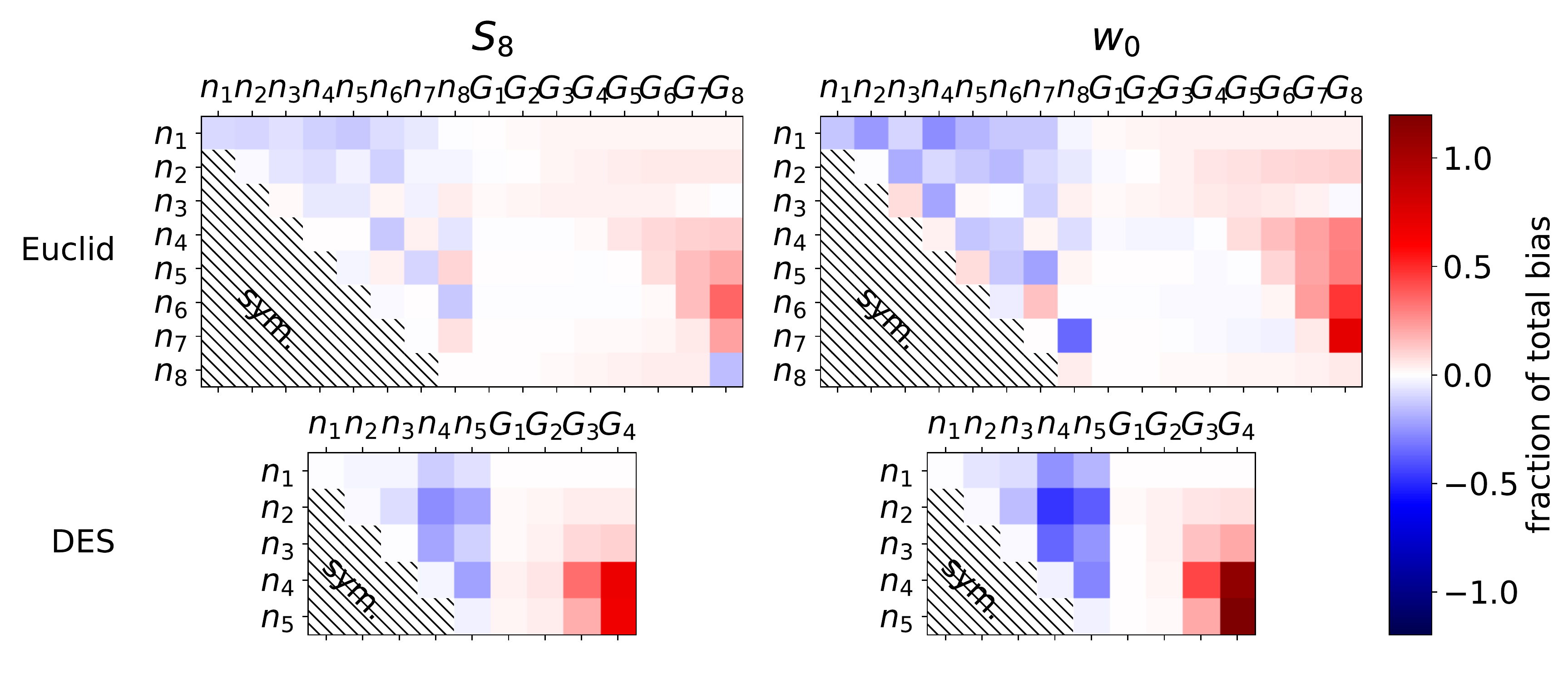}
\caption{Contributions to the total bias on $S_8$ and $w_0$ for both the Stage-IV (Euclid-like) and
the Stage-III (DES-like) survey we are considering. Note that the results for the two different
DES-like setups (``Y1'' and ``Full'') would look identical in our model, since they differ only in their covered sky-fraction. This increase in survey coverage does not affect the biases. Note also that for the results presented elsewhere in this paper, the cross-correlations in clustering are removed from the DES-like data vector, as in \citep{DES_3x2pt_Abbot}; we show them here for illustration and comparison to the Stage-IV survey.}\label{fig:individualpairbiases}
\end{figure*}

This curious finding may be better understood by considering Fig.~\ref{fig:individualpairbiases}. In
this figure, we plot the relative contributions to the total bias by all relevant redshift bin
combinations, both for the Euclid and the DES model, on the parameters $S_8$ and $w_0$. Similar
plots are obtained for other parameters as well but not included here. We also note that we found
the qualitative results discussed in this section to be robust against changes in the prescription
for the non-linear scale cuts. Note that the shear-shear combinations do not have any magnification
contribution at first order, and that the clustering-clustering part of the matrices plotted is
necessarily symmetric. We see a pattern that is roughly consistent both between different
cosmological parameters as well as between different survey models. There are large contributions to
the bias from correlations of high-redshift shear with high-redshift clustering bins. The
contributions to the bias from the galaxy-galaxy lensing part of the data are predominantly in the
same direction, while the contributions from the clustering-clustering part are in the opposite
direction. These findings agree with the results of \cite{Lorenz18} (c.f. their Table~1). The overall bias is therefore the result of partial cancellation from the various combinations in the data vector, which can result in an increase in the sensitivity to magnification with the removal or elements of the data vector, as noted earlier. This is emphasised by the scaling of the colour-bar in Fig. \ref{fig:individualpairbiases}, which shows that the largest individual contribution is larger than the total summed bias.

Interestingly, we find that redshift bin combinations whose power spectra are entirely dominated by the magnification signal (e.g., $G_1$-$n_8$, where the shear sample is at lower redshift than the lens sample, and widely separated clustering correlations such as $n_1$-$n_8$, compare Fig.~\ref{fig:GGL-power-spectra}) do not necessarily contribute large fractions to the total bias. Instead, for both survey models, the largest contributions to the parameter bias comes from the galaxy-galaxy lensing data vector, and in particular the correlations between the high redshift background (shear) and the lower redshift foreground (lens) samples, which are dominated by the traditional (non-magnification) galaxy-galaxy lensing signal.

This seemingly counter-intuitive result may be understood by considering how an additional shift in the data propagates to a parameter bias. In this experiment, we ask by how much must the model parameters be shifted to account for an un-modelled contribution to the data, in this case coming from the inclusion of the magnification bias in the data. The magnitude of parameter bias that one gets from a systematic shift in a fixed parameter results not only due the absolute difference in the data vector, but also in the sensitivity of the remaining (incomplete) model which is used to fit the shifted data. In the formalism of Eq. \ref{eqn:FMBiasZ}, where the correlation between cosmological parameters and the shift in the data vector due to magnification (given by $\sum_J\Delta_{\alpha} C^I ({\rm Cov}^{-1})^{I,J}\partial_{\tau}C^{J}$, where the sum over $\ell$-modes is implied) is small compared to the sensitivity of the model to the data without a magnification term (given by $F$), the model is unable to mimic the magnification contribution and the absolute bias in cosmological parameters is small.

To understand this better, consider the example of a fit to the position-position power spectrum for two widely separated redshift bins, with no overlap in their distributions. In this case, the data should not contain any power due to intrinsic clustering (which would occur only if they experience the same gravitational potential). The data will include a non-zero contribution due to the magnification (including the correlation due to the magnification of the background by the potential well in which the foreground resides, and the correlation which results from lensing by all large-scale structure forward of the foreground). In this case, the magnification effect is therefore the only contribution to the data. Where the magnification effect is ignored in the model, we attempt to fit this data with the intrinsic clustering signal. As there is no overlap in the sample redshift distributions, the model is zero for all parameter choices, and therefore the data is equally badly fit by the model conditioned on any cosmological parameter. As a result, the contribution to the likelihood of this particular bin combination is flat, and therefore it does not contribute to a parameter bias. This is true irrespective of the size of the magnification signal, and results solely from the insensitivity of the model being fit to variations in cosmological parameters.

It is worth noting that for the example we have described here (and in our models for both DES- and
Euclid-like surveys), we have implicitly assumed that the redshift distributions for the foreground
and background are known, or have strong prior information. If the main aspects of the distribution
(e.g. the mean, variance or overall shape) are inferred from the data, the magnification
contribution to the data could be partially mimicked by shifting the model redshift distributions or
increasing their tails, so that there is an increased overlap in the model redshift distributions in
the position-position terms, or the redshift distribution and the lensing kernel in the
position-shear terms, effectively ``biasing'' the model redshift distributions. This, in turn can
impact the accuracy of the inferred cosmological parameters due to correlation between these
parameters and those which describe the model redshift distributions. Due to the wide width of the
lensing kernel, it can be expected that the position-shear and shear-shear correlations will be less
sensitive to the bias in the model redshift distributions due to magnification than the clustering
part, so that a certain amount of `self-calibration' of the redshift distribution models will occur
in the 3x2pt analysis. For the DES-Y1-like survey model, we tested the sensitivity to the redshift
distributions by simply perturbing the mean of the highest redshift bin by an amount equal to the prior width applied to it in the analysis of \cite{DES_3x2pt_Abbot}, and found that for most cosmological parameters the resulting bias is comparable in size to that resulting from magnification. We therefore conclude that the investigation into the self-calibration of the redshift distribution model and impact on inferred cosmological parameters merits further study.

Putting the findings derived from Fig.~\ref{fig:individualpairbiases} together, a clearer picture explaining the counter-intuitive results listed in Tab.~\ref{tab:Euclidcuttingbiases} emerges:
\begin{itemize}
\item First, the largest biases are not necessarily contributed by the redshift bin combinations whose power spectra have the largest magnification contribution. We therefore see that the removal (or non-inclusion) of redshift bin combinations dominated by the magnification signal such as the clustering cross-correlations and the low redshift shear, high redshift lens galaxy-galaxy lensing signal is not enough to sufficiently remove the bias due to the magnification contribution (although they will contribute to a bad goodness-of-fit). As a result, future analyses must explicitly measure and model the magnification signal to avoid such bias, even if these combinations are not included in the data vector.
\item Second, different redshift bin combinations to the data vector can contribute biases of different signs, so that partial cancellation occurs between different elements of the data vector. Moreover, we see that typically the clustering cross-correlations induce a bias of opposite sign to the galaxy-galaxy-lensing parts, so that the removal of only the clustering cross-correlations can increase the sensitivity to magnification, as demonstrated in \cite{Duncanetal2014}.
\end{itemize}

We emphasise that the information given in Fig.~\ref{fig:individualpairbiases} cannot possibly be used to construct a redshift bin combination cutting strategy that would reduce the magnification bias while keeping degradation in the FoM reasonable. Besides the inherent limitations of the analysis presented here, including the Fisher approximation and the fact that we are working in $\ell$-space, the complex structure of the contributions to the total bias would render such an approach impossible. This point is further illustrated in Appendix~\ref{app:contrivedcutting}, where we consider how the FoM and bias change in the contrived case where the redshift bin combinations are cut in order of the largest contribution to the bias according to Fig.~\ref{fig:individualpairbiases}. The results show a non-trivial behaviour which demonstrates that, even in this case, the relative bias does not necessarily decrease even though the FoM is reduced due to the removal of cosmological information. This example further underlines the conclusion that no simple mitigation strategy exists that solely relies on the removal of redshift bin combinations.

\subsection{Sensitivity to scale cuts}
\label{sec:scalecuts}
As detailed in Section \ref{sec:SurveyModel_Euclid}, for the Euclid-like survey we apply non-linear
cuts in Fourier space by cutting scales beyond which the variation of the matter density within that
scale (parameterised as $\sigma_R$) exceeds a given value ($\sigma_R = 0.5$ in our fiducial model). In this section, we study the dependence of constraining power (encoded in the FoM) and the impact of magnification bias on the choice of $\sigma_R$. Whereas \cite{Duncanetal2014} only considered the improvement in the FoM as more non-linear scales are included (i.e., $\sigma_R$ is increased), we show here that the effect of magnification bias is strongly dependent on the scale cuts.

\begin{figure}
\includegraphics[width=\linewidth]{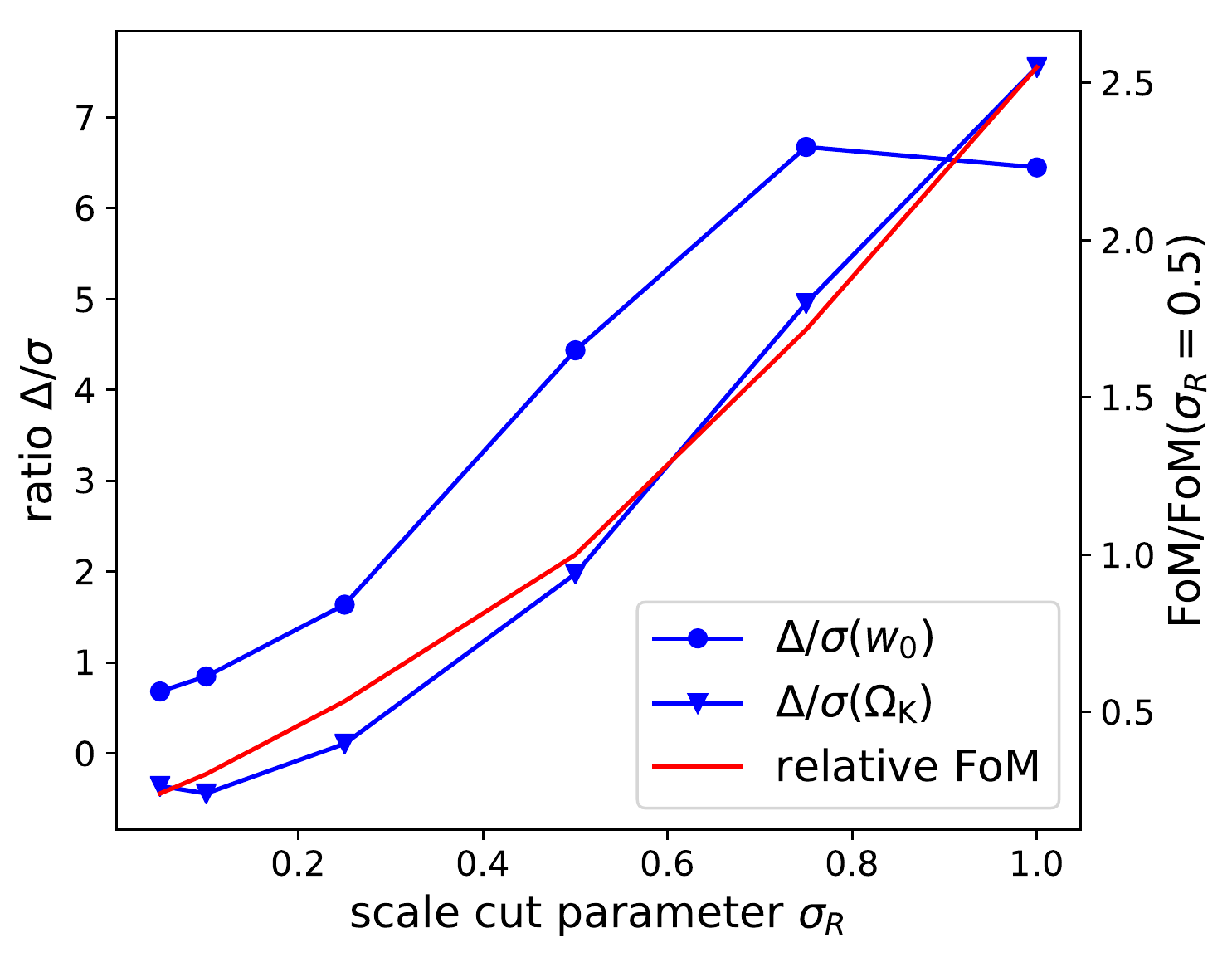}
\caption{The relative parameter bias $\Delta/\sigma$ as a function of the scale cut parameter $\sigma_R$, for the two parameters $w_0$ (circles) and $\Omega_{\rm K}$ (inverted triangles), in blue. The red line also shows the figure of merit relative to its value for $\sigma_R=0.5$. This forecast is for the Euclid-like survey. As $\sigma_R$ increases, increasingly smaller scales are included in the analysis. We point out that the increase in $\Delta/\sigma$ with increasing $\sigma_R$ is not solely driven by the increasing constraining power, i.e. decreasing $\sigma$, but $\Delta$ alone increases as well.}
\label{fig:scalecuts}
\end{figure}
In Fig.~\ref{fig:scalecuts} we plot both the relative bias for two parameters and the full FoM as a function of the scale cut parameter $\sigma_R$. The bias is highly sensitive to the scale cuts. As more conservative cuts are applied, we find that the biases on $w_0$ and $\Omega_k$ decrease so that, for sufficiently strict cuts on the non-linear scales, the parameter bias can always become smaller than the statistical uncertainties. We see, however, that this comes at significant cost to the constraining power of the survey.

The change in relative bias is not solely driven by the change in the parameter uncertainty -- in fact, between $\sigma_R = 0.5$ and $\sigma_R = 0.1$ we find that the absolute bias decreases by a factor of $2$ and $3$ for $w_0$ and $\Omega_{\rm K}$ respectively. The reduction of the sensitivity to the magnification therefore comes from a combination of decreasing statistical power and decreasing absolute bias. Although we have chosen to plot the behaviour of the relative bias for only two parameters, the same qualitative behaviour is observed for all cosmological parameters considered in this work, with the exception of $\Omega_{\rm B}$, for which the relative bias fluctuates around $20\,\%$ regardless of scale cuts.

\section{Conclusions}\label{sec:Conclusions}
Upcoming galaxy surveys will measure weak lensing shear and number density power spectra with unprecedented precision. Increased constraining power, however, needs to be matched with equally accurate control for systematic errors. Here, we have focused on a 3$\times$2-point analysis, utilising shear and position auto- and cross-power spectra. For such an analysis, previous work identified magnification bias as a severe problem, showing that a Stage-IV survey would be substantially biased if magnification is not controlled for. In this work, we disentangle how the total magnification bias on cosmological parameters arises as a combination of different regimes in the data vector. We illustrate our findings on two survey models, namely a current Stage III survey (modelled after DES-Y1 results) and a future Stage IV survey (modelling a Euclid-like experiment).

These survey models contain several simplifying assumptions; for example, we assume well-known redshift distributions without catastrophic outliers; magnification bias sourced only from a hard magnitude cuts at the faint end of the survey with no selection based on size; and a constant $\alpha = 0.7$ for the Stage IV survey model (by contrast the DES survey model uses a redshift-varying $\alpha$ inferred from the Y1 data)

As a first result, we compute the total biases on cosmological parameters if the magnification effect is not accounted for in the model, but is present in the clustering and galaxy-galaxy-lensing data. Our results are consistent with previous work as detailed in \cite{Duncanetal2014, Lorenz18, Cardona16}. We find that parameter biases are typically comparable to or smaller than statistical uncertainty for the DES-Y1 application. However, we find that  a simplified model of the full DES survey (where the survey area is simply boosted over the Y1 model) and the Euclid-like survey both display significant parameter biases, suggesting that current surveys are reaching the point where sensitivity to the magnification means it needs to be included in the model which is fit to the data.

We extend the bias formalism to consider the contribution to the overall bias coming from each of the different redshift bin correlations which enter the data vector. For both surveys, and consistent between different cosmological parameters, we find a relatively complex structure in the contributions arising from different redshift bin correlations. Perhaps contrary to na\"{i}ve intuition, the dominant contributions to the total bias do not arise from the redshift bin correlations whose power spectra are dominated by the magnification signal. We have demonstrated that the removal of these contributions from the data vector does not reduce the sensitivity to magnification, and can instead increase the significance of any parameter bias. The removal of magnification-dominated redshift correlations (such as redshift bin cross-correlations in the clustering, and low-redshift shear cross correlated with high-redshift lens samples in the galaxy-galaxy-lensing) does not guarantee a reduced sensitivity to the magnification in the fit model. Moreover, in the contrived case where knowledge on the sensitivity of the model conditioned on each element of the data vector is known a-priori, we found that the removal of the cross-correlations with the largest absolute contribution to the total bias would not necessarily reduce the overall bias, due to partial cancellation of terms across all components of the data vector. 

We further demonstrated that the sensitivity to the magnification can be reduced by the application of stricter cuts on small scales, however we find that this comes at the cost of a significant increase in parameter uncertainty. The conclusion is therefore, that it will not be straightforward to avoid the impact of magnification in near future surveys (e.g. through bin-pair or scale cuts), and it must be accurately modelled.

It has to be emphasised that the simplified models we used in this work cannot be considered a complete description of how magnification bias can manifest itself in an actual analysis; rather, a wide range of systematic effects needs to be considered. In this work, we considered the case where the magnification bias was sourced only from the application of a hard cut at the faint limit of the survey, and ignored potential contributions from bright magnitudes and size cuts, which could increase the sensitivity to the magnification bias in the type of analysis considered. For example, we did not include the effect of magnification on the shear sample, through selection and through correlations between magnification, source signal-to-noise and inferred shear. Whilst \cite{Duncanetal2014, Lorenz18} suggest that the magnification strength $\alpha$ must be known to a few percent to ensure that the bias due to the magnification contribution is sub-dominant to statistical uncertainty in Stage-IV experiments, in practice, the accurate simultaneous inference of the magnification contribution is complicated by a range of systematics which can be survey or sample dependent, such as:
\begin{itemize}
  \item Survey selection function: The parameterisation of the magnification strength using Eq.
  \ref{eqn:alpha_m} is exact only in the case where a hard selection has been applied on the basis
  of source flux or magnitude, or where the galaxy number counts are well known as a function of magnitude. Sample selection on observables correlated to source magnitude, such as galaxy morphology (including ellipticity), signal-to-noise ratio or colours can induce a smooth rather than hard selection function on the source number counts. Determination of $\alpha$ on the observationally incomplete sample will induce a bias on the determination of the magnification strength. For an example of this, see Figure 2 of \cite{HH_MagBiases}, which shows that both the determination of magnification through $\alpha$ can be biased due to the weak lensing approximation, and further biased due to the selection function. In such a case, one can either further select the source sample to be fully complete; by determining $\alpha$ from a deeper survey, which is complete at the limiting magnitude of interest\footnote{Note that the magnitude where the deeper survey is complete becomes the limiting magnitude, which must be applied as a hard cut to the data of interest if the selection function is not zero by that magnitude.} (if the weak lensing approximation is sufficient) \citep{HH_MagBiases}; or by accounting for the selection function as part of the inference process using a model of the intrinsic number counts \cite[see e.g.][ for an example application in the process of Bayesian inference in direct size and magnitude measurements]{Alsing2015, Duncan2016}.
  \item Dust attenuation: Reddening of the background sample due to dust in the foreground can cause a depletion of background sources which gives rise to correlations which can mimic the magnification \cite[e.g.][]{HH_Submillimetre}. The level of dust extinction can be simultaneously inferred from the data by fitting as a free parameter \cite[e.g.][]{HH_Submillimetre}, or using colour excess measurements \cite[e.g.][]{Menard_Extinction_ColorExcess}. However, the presence of photometric noise can induce chromaticity in the magnitude shift across bands in the presence of a colour selection, which can induce bias in the inferred extinction which may be tomography dependent \cite[][]{HH_MagBiases}.
  \item Spatial variation in magnitude zero-point or survey depth causes local changes in the number density which mimic the magnification signal \cite[e.g.][]{Morrison_Systematics_Depth}.
  \item Sample contamination, such as star-galaxy separation and subsequent local variation of size and magnitude cuts.
  \item Background sample detection and selection: obscuration of the background by the foreground
  galaxy reduces the number of detected sources as a function of separation \cite[see e.g.][for an
  example in application to galaxy clusters]{Tudorica_Sparcs, Umestsu_11_Cluster}. Additionally, the
  magnification of the background can induce correlations between the shear and magnification fields
  through the change in the measured properties of the foreground or source. Note that in
  \cite{Martinet_Euclid_Background} it was shown that the impact of the magnification of an unresolved background population of galaxies has negligible impact on the multiplicative bias of the shear of the foreground. Further, selection of the background shear sample can induce additional correlations in the shear and galaxy-galaxy lensing signal due to magnification \cite[][]{SchmidtRozo2009,Liu2014}.
  \item Accurate determination of the redshift distributions for photometric data, especially the tails of the distribution.
\end{itemize}
The investigation of the sensitivity of future analyses to these effects is therefore strongly encouraged.

Regardless of these systematics, however, this work suggests that there is no simple cutting scheme on the data vector which will adequately reduce the sensitivity to magnification bias. Moreover, the application of such a cutting scheme reduces the constraining power of the analysis as by nature non-magnification contributions are also cut. As a result we recommend that the magnification contribution be modelled and fit in forthcoming surveys.

\section*{Acknowledgements}
All authors should be considered equal contributors to this paper. LT acknowledges support by the Studienstiftung des Deutschen Volkes. CAJD acknowledges support from the Beecroft Trust. DA acknowledges support from the Beecroft Trust and from STFC through an Ernest Rutherford Fellowship, grant reference ST/P004474/1. Research at Perimeter Institute is supported in part by the Government of Canada through the Department of Innovation, Science and Economic Development Canada and by the Province of Ontario through the Ministry of Economic Development, Job Creation and Trade. This project used public archival data from the Dark Energy Survey (DES). Funding for the DES Projects has been provided by the U.S. Department of Energy, the U.S. National Science Foundation, the Ministry of Science and Education of Spain, the Science and Technology FacilitiesCouncil of the United Kingdom, the Higher Education Funding Council for England, the National Center for Supercomputing Applications at the University of Illinois at Urbana-Champaign, the Kavli Institute of Cosmological Physics at the University of Chicago, the Center for Cosmology and Astro-Particle Physics at the Ohio State University, the Mitchell Institute for Fundamental Physics and Astronomy at Texas A\&M University, Financiadora de Estudos e Projetos, Funda{\c c}{\~a}o Carlos Chagas Filho de Amparo {\`a} Pesquisa do Estado do Rio de Janeiro, Conselho Nacional de Desenvolvimento Cient{\'i}fico e Tecnol{\'o}gico and the Minist{\'e}rio da Ci{\^e}ncia, Tecnologia e Inova{\c c}{\~a}o, the Deutsche Forschungsgemeinschaft, and the Collaborating Institutions in the Dark Energy Survey. The Collaborating Institutions are Argonne National Laboratory, the University of California at Santa Cruz, the University of Cambridge, Centro de Investigaciones Energ{\'e}ticas, Medioambientales y Tecnol{\'o}gicas-Madrid, the University of Chicago, University College London, the DES-Brazil Consortium, the University of Edinburgh, the Eidgen{\"o}ssische Technische Hochschule (ETH) Z{\"u}rich,  Fermi National Accelerator Laboratory, the University of Illinois at Urbana-Champaign, the Institut de Ci{\`e}ncies de l'Espai (IEEC/CSIC), the Institut de F{\'i}sica d'Altes Energies, Lawrence Berkeley National Laboratory, the Ludwig-Maximilians Universit{\"a}t M{\"u}nchen and the associated Excellence Cluster Universe, the University of Michigan, the National Optical Astronomy Observatory, the University of Nottingham, The Ohio State University, the OzDES Membership Consortium, the University of Pennsylvania, the University of Portsmouth, SLAC National Accelerator Laboratory, Stanford University, the University of Sussex, and Texas A\&M University. Based in part on observations at Cerro Tololo Inter-American Observatory, National Optical Astronomy Observatory, which is operated by the Association of Universities for Research in Astronomy (AURA) under a cooperative agreement with the National Science Foundation.


\appendix

\section{Measuring magnification strength in DES-Y1}
\label{app:DESalpha}

We use the public DES-Y1 data to determine the magnification strength for the selection of galaxies used in the clustering analysis, and as the lenses in the galaxy-galaxy lensing contribution. These are redMaGiC selected galaxies, which selects based on a minimum luminosity threshold, and based on their goodness-of-fit to a red sequence template calibrated with redMaPPer on an overlapping spectroscopic sample \cite[see e.g.][for  further details.]{DES_2pt_Clustering_ElvinPoole}. The first three redshift bins are determined with a low-luminosity threshold of $L > 0.5L_{\star}$ (denoted the High Density sample), whilst the two highest redshift bins are selected with brighter luminosity cuts of $L>L_{\star}$ and $L>1.5L_{\star}$ (denoted the High and Higher Luminosity samples respectively). As a luminosity thresholded sample, we therefore derive the magnification strength using equation \ref{eqn:alpha_m}, based on the z-band absolute rather than apparent magnitude.

\begin{figure}
  \centering
  \includegraphics[width = 0.5\textwidth]{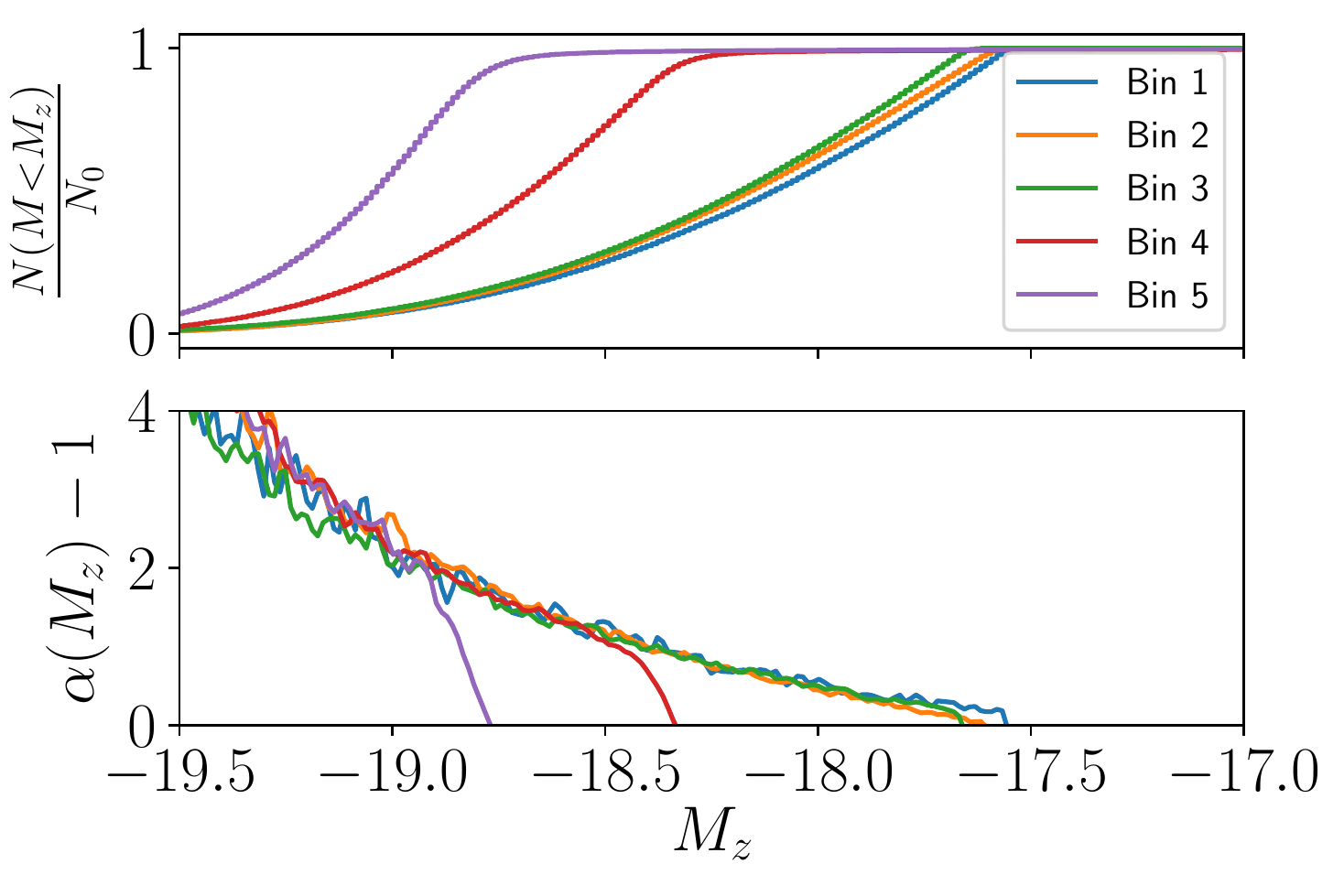}
  \caption{Top: Cumulative probability density for the number counts of sources in DES-Y1
  photometric clustering sample, as a function of redshift bin. Bottom: Magnification strength,
  shown as $\alpha - 1$ for the clustering sample for DES-Y1, as a function of z-band absolute
  magnitude. This is also the lens sample in the galaxy-galaxy lensing contribution to the data
  vector. At low magnitudes, the magnification strength is similar for all bins, whilst they diverge
  strongly for high redshift bins at fainter magnitudes due to selection.} \label{fig:DESY1_alpha}
\end{figure}

Figure \ref{fig:DESY1_alpha} shows the normalised cumulative distribution, and magnification strength for all redshift bins. We see in the top panel that the cumulative distributions for the lowest three redshift bins are very similar down to $M_z \sim -17.5$, where they show a rapid flattening consistent with a sharp cut in the luminosity function. The two highest redshift bins are fully incomplete by $M_z \sim -18.4$ and $M_z \sim -18.7$, consistent with the brighter luminosity threshold applied. In contrast to the lower three redshift bins, the highest redshift bins show a more gradual flattening of the distribution, which indicates a smooth, non-sharp selection function at faint magnitudes. Since $\alpha-1$ is only a good approximation to the magnification strength to the limit where the survey is complete, it can be expected that the actual change in number density due to magnification in these samples will differ from that predicted by equation \ref{eqn:N_variation_mag} \citep{HH_MagBiases}. In such a case, the magnification should be inferred from deeper data which is complete to the magnitude in which the selection on the data of interest goes to zero. In application to the DES-Y1 data-set, such a correction will be largest for the highest redshift bins, which demonstrate a slower rate of incompleteness with magnitude, and where the underlying luminosity function is steeper at the luminosity limit considered. For the purposes of this forecast, it is expected that this approximation is sufficient, however for a full analysis it may be necessary to understand the impact of this smooth selection function in more detail.

The bottom panel of Figure \ref{fig:DESY1_alpha} shows the derived magnification strength for these samples. At bright magnitudes we see that the magnification strength is approximately the same for all redshift bins, consistent with little change in the underlying luminosity function at the bright end with redshift for this data-set.  The two highest redshift bins become fully incomplete (consistent with $\alpha = -1$) at their limiting magnitudes, where the slope of the luminosity function is relatively larger. Consequently, we find that the highest redshift bins have a larger magnification strength due to the fact that they select on the underlying luminosity function with a brighter magnitude limit. For the purpose of our predictions, we choose the values of $\alpha$ at the limiting magnitude for each redshift bin, giving $\alpha(z_i) = \{ 2.0, 1.5, 0.3, 0.3, 0.3\}$.

\section{Contrived bias mitigation strategy}\label{app:contrivedcutting}
\begin{figure}
\includegraphics[width=0.5\textwidth]{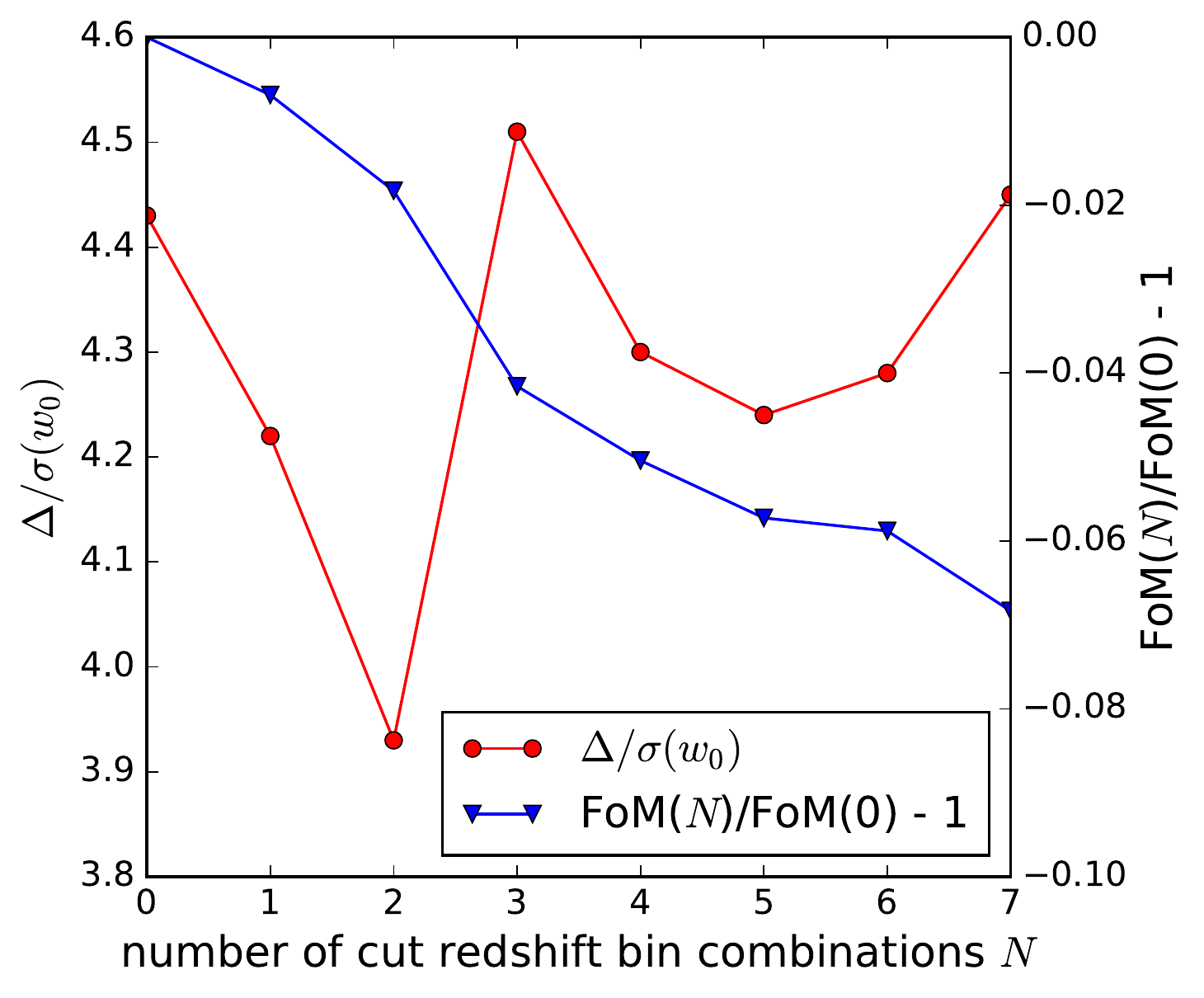}
\caption{Behaviour of relative bias $\Delta/\sigma$ on the parameter $w_0$, as well as the all-parameter figure of merit (FoM), as increasingly more redshift bin correlations are excluded from the analysis of our Euclid survey model. The order of cutting is determined by the absolute values of the matrix elements in the upper right panel of Fig.~\ref{fig:individualpairbiases} (i.e., for $N=1$ the cell n7-G8 is removed, for $N=2$ in addition n6-G8 is cut, etc.)} \label{fig:Euclidrankedcuts}
\end{figure}

As alluded to in the main text, a redshift bin combination cutting scheme based on the information given in Fig.~\ref{fig:individualpairbiases} is not only highly contrived, but also not successful even within the approximations considered in this work. This point is illustrated in Fig.~\ref{fig:Euclidrankedcuts}. We use the information contained in Fig.~\ref{fig:individualpairbiases} in order to devise a possible redshift bin correlations cutting scheme that could perhaps be expected to reduce the bias on $w_0$. As can be seen in the figure, the fractional bias is actually not necessarily decreasing as redshift bin correlations that are large contributors (by absolute value) to the total bias are cut out. This can be explained by both the partial cancellations (because different large contributions to the bias do not always have the same sign), as well as the changing Fisher matrix. Thus, although the information loss is relatively minor (when seven correlations have been removed, the FoM has degraded by $\sim 7\%$), a cutting scheme constructed as described appears not to yield the desired decrease in bias. Of course, one could refine the cutting strategy by taking not only the absolute value but also the sign of the individual contributions into account, but such an approach appears very contrived (one would keep certain biases in the expectation that they cancel others). We also point out that we have concentrated on a single parameter here ($w_0$), however, if the analysis were to be extended to multiple parameters a unique bias ranking of redshift bin correlations would be impossible.

\bsp
\label{lastpage}
\end{document}